\documentclass[journal]{vgtc}                

\usepackage{microtype}                 
\PassOptionsToPackage{warn}{textcomp}  
\usepackage{textcomp}                  
\usepackage{amsmath}
\usepackage{graphicx}
\usepackage[]{algorithm2e}
\usepackage{mathtools}
\usepackage{amssymb}
\usepackage{threeparttable}
\usepackage{caption}
\usepackage{times}
\usepackage{cite}                      
\usepackage{tabu}                      
\usepackage{booktabs}                  

\usepackage{adjustbox} 
\usepackage{xcolor}

\newcommand{\beginsupplement}{%
        \setcounter{table}{0}
        \renewcommand{\thetable}{S\arabic{table}}%
        \setcounter{figure}{0}
        \renewcommand{\thefigure}{S\arabic{figure}}%
        \setcounter{section}{0}
        \renewcommand{\thesection}{S\arabic{section}}  %
     }

\onlineid{1095}

\vgtccategory{Research}
\vgtcpapertype{Algorithm / Technique}

\title{VC-Net: Deep Volume-Composition Networks for Segmentation and Visualization of Highly Sparse and Noisy Image Data}

\author{Yifan Wang, Guoli Yan, Haikuan Zhu, Sagar Buch, Ying Wang, Ewart Mark Haacke, Jing Hua, and Zichun Zhong}
\authorfooter{
\item
Y. Wang, G. Yan, H. Zhu, J. Hua, and Z. Zhong are with the Department of Computer Science, Wayne State University, Detroit, MI 48202. E-mail: \{yifan.wang2,guoliyan,hkzhu,jinghua,zichunzhong\}@wayne.edu.
\item
S. Buch, Y. Wang, and E. M. Haacke are with the Department of Radiology, Wayne State University, Detroit, MI 48201. E-mail: \{sagarbuchmri,neuying\}@gmail.com, nmrimaging@aol.com.
}

\shortauthortitle{Wang \MakeLowercase{\textit{et al.}}: VC-Net: Deep Volume-Composition Networks for Segmentation and Visualization}

\abstract{The fundamental motivation of the proposed work is to present a new visualization-guided computing paradigm to combine direct 3D volume processing and volume rendered clues for effective 3D exploration. For example, extracting and visualizing microstructures \emph{in-vivo} have been a long-standing challenging problem. However, due to the high sparseness and noisiness in cerebrovasculature data as well as highly complex geometry and topology variations of micro vessels, it is still extremely challenging to extract the complete 3D vessel structure and visualize it in 3D with high fidelity. In this paper, we present an end-to-end deep learning method, \emph{VC-Net}, for robust extraction of 3D microvascular structure through embedding the image composition, generated by maximum intensity projection (MIP), into the 3D volumetric image learning process to enhance the overall performance. The core novelty is to automatically leverage the volume visualization technique (e.g., MIP -- a volume rendering scheme for 3D volume images) to enhance the 3D data exploration at the deep learning level. The MIP embedding features can enhance the local vessel signal (through canceling out the noise) and adapt to the geometric variability and scalability of vessels, which is of great importance in microvascular tracking. A multi-stream convolutional neural network (CNN) framework is proposed to effectively learn the 3D volume and 2D MIP feature vectors, respectively, and then explore their inter-dependencies in a joint volume-composition embedding space by unprojecting the 2D feature vectors into the 3D volume embedding space. It is noted that the proposed framework can better capture the small / micro vessels and improve the vessel connectivity. To our knowledge, this is the first time that a deep learning framework is proposed to construct a joint convolutional embedding space, where the computed vessel probabilities from volume rendering based 2D projection and 3D volume can be explored and integrated synergistically. Experimental results are evaluated and compared with the traditional 3D vessel segmentation methods and the state-of-the-art in deep learning, by using extensive public and real patient (micro-)cerebrovascular image datasets. The application of this accurate segmentation and visualization of sparse and complicated 3D microvascular structure facilitated by our method demonstrates the potential in a powerful MR arteriogram and venogram diagnosis of vascular disease.} 


\CCScatlist{ 
 \CCScat{K.6.1}{Management of Computing and Information Systems}%
{Project and People Management}{Life Cycle};
 \CCScat{K.7.m}{The Computing Profession}{Miscellaneous}{Ethics}
}

\vgtcinsertpkg

\begin{document}


\firstsection{Introduction}

\maketitle
\label{sec:intro}
Nowadays, there is a pressing need for better visualizing and understanding microstructures in the raw and wild datasets. For instance, the acquisition of the \emph{in-vivo} micro-level 3D vasculature from image data is a grand challenge. The notorious difficulties of microvascular data analytics lie in \emph{high sparseness} of vessel data in a large-sized 3D volume, e.g., the scattered vessel fragments in the angiographic data against the otherwise encompassing white and grey matter (as well as background and noise); \emph{high noisiness}, such as low signal-to-noise ratio (SNR), e.g., about 10:1 in cerebrovascular images; tininess of micro-level vessels, e.g., the diameter of the micro-level vessels in images is merely 1$\sim$2 voxels (e.g., 50$\sim$100 microns); and sophisticated vessel geometry and topology variations, e.g., local ``crossing'', ``kissing'', or ``tortuous'' vessel structures, etc. Currently, for such complex 3D micro-level data, it would be impossible from a timing perspective for clinicians to review all this data manually and label abnormalities slice by slice. The 3D structural / contextual information and quantitative metrics are still missing, although maximum intensity projection (MIP)~\cite{napel1992ct}, a widely-used approach for qualitatively visualizing and analyzing the 3D vasculature, has been employed to enhance the local vessel signal, allowing for geometric variability and scalability. The labor-intensive, time-consuming, and 3D global / contextual information-missing nature of the procedure makes it very challenging to fully take advantage of the large number of 3D datasets (images and shapes) available for reference and comparison, and reach more informed and accurate decisions.

In recent decades, the automatic model-driven vessel extraction and segmentation approaches have been proposed, such as multiscale filtering~\cite{frangi1998multiscale}, region growing techniques~\cite{martinez1999retinal}, active contours~\cite{nain2004vessel}, geometric flow~\cite{descoteaux2008geometric}, level-set approach~\cite{forkert20133d}, nonlinear subtraction (NLS) method~\cite{ye2013noncontrast}, template-based predictor-corrector algorithm~\cite{govyadinov2018robust}, etc. However, these approaches are easily overwhelmed by tons of low-level handcrafted features and complicated manual parameter adjustment to overcome aforementioned difficulties and subject variations.

Recently, data-driven approaches have been proposed to robustly investigate the correlations between different objects / instances without relying on hard-coded metrics. In medical image visualization and processing, several deep learning based methods have been proposed to extract vessels from 2D retinal images, such as DeepVessel~\cite{fu2016deepvessel}, multi-level deep supervised networks~\cite{mo2017multi}, deep neural network (DNN)-based method~\cite{liskowski2016segmenting}, unified convolutional neural network (CNN) and graph neural network (GNN)~\cite{shin2019deep}, etc. These methods can perform 2D vessel segmentation tasks well, but are far from satisfactory on 3D vessel scenario. There are still very few dedicated deep learning architectures for 3D vessel segmentation, such as Uception~\cite{sanchesa2019cerebrovascular}, DeepVesselNet~\cite{tetteh2018deepvesselnet}, and VesselNet~\cite{kitrungrotsakul2019vesselnet}, etc. Existing methods do not consider to use the visualization techniques in the 3D vessel extraction and are not specifically designed for solving the aforementioned challenges in 3D micro-cerebrovascular segmentation.


The fundamental motivation of the proposed work is to present a new \emph{visualization-guided computing paradigm} to combine direct 3D volume processing and volume rendered clues for effective 3D exploration. In order to fill the gap in the high-fidelity 3D micro-cerebrovascular segmentation and visualization for the medical data \emph{in-vivo}, we present a DNN method, \emph{VC-Net}, for robustly extracting sparse microvascular structures through embedding the 2D image slice composition by MIP into the 3D volumetric image learning process to enhance the overall performance on 3D vasculature segmentation. The core \emph{novelty} is to automatically leverage the volume visualization technique (e.g., MIP -- a volume rendering technique for 3D volume images) to enhance the qualitative 3D data exploration, especially for 3D \emph{in-vivo} segmentation and visualization, at the deep learning level. It is noted that the proposed framework can better capture the micro vessels and improve the vessel connectivity. The key motivation of our network is to integrate the trustworthy auxiliary from learned 2D MIP features into the 3D volume segmentation and visualization network, instead of using more complicated networks empirically. Experimental results are evaluated and compared with the traditional 3D vessel segmentation methods and the state-of-the-art in deep learning, using extensive public and real patient (micro-)cerebrovascular image datasets. The key \emph{contributions} of our work are as follows:\vspace{-2mm}
\begin{itemize}
 \item It proposes an effective end-to-end deep learning method to segment and visualize high-fidelity 3D sparse microvascular structure with complicated geometry and topology variations from volumetric images with significant noise.\vspace{-2mm}
  \item A multi-stream CNN framework is designed to effectively learn the feature vectors of 3D raw volume and multislice composited 2D MIP (volume rendering), respectively, and explore inter-dependencies between 3D and 2D embedded features in a joint volume-composition embedding space by unprojecting (inverse volume rendering) the 2D features, learned from MIP, into the 3D volume embedding space.\vspace{-2mm}
 \item To our knowledge, this is the first time that a deep learning framework is proposed to construct such a joint convolutional embedding space, where the computed joint vessel probabilities from 2D projection and 3D volume can be integrated synergistically.\vspace{-2mm}
  \item The application and experiments on the accurate \emph{in-vivo} segmentation and visualization of sparse and complicated 3D microvascular structure facilitated by our method demonstrate the potential in a novel and powerful MR arteriogram and venogram (MRAV) diagnosis of vascular disease.\vspace{-0mm}
\end{itemize}

\section{Related Work}
\label{sec:relatedwork}
In this section, we review most related work on 2D / 3D vessel extraction and segmentation in visualization and medical imaging domains.\vspace{-1mm}
\vspace{-1mm}
\subsection{Model-Driven Vessel Extraction and Segmentation}
Traditionally, doctors have to manually segment each image slice to obtain accurate vessel structures, which is extremely tedious and time-consuming. Therefore, it is important to develop automatic vessel segmentation methods. For instance, Wilson and Noble~\cite{wilson1997segmentation} introduced a mixture distribution for the data, motivated by a physical model of blood flow, that is used in a two-stage segmentation algorithm with a statistical classifier and structural criteria. Chung and Noble~\cite{chung1999statistical} presented an extended version of the previous 3D cerebral vessel segmentation algorithm~\cite{wilson1997segmentation}, and introduced a Rician distribution for background noise modeling and used a modified expectation-maximization (EM) algorithm for the parameter estimation procedure. Frangi et al.~\cite{frangi1998multiscale} developed a vessel enhancement filter by computing the multiscale second order local structure of an image (i.e., Hessian). A vesselness measure is obtained on the basis of all eigenvalues of the Hessian. Based on multiscale filtering method~\cite{frangi1998multiscale}, Descoteaux et al.~\cite{descoteaux2008geometric} developed a novel geometric flow for segmenting vasculature in proton-density images, which can also be applied to the cases of magnetic resonance angiography (MRA) or MRI data. Mart{\'\i}nez-P{\'e}rez et al.~\cite{martinez1999retinal} presented a retinal blood vessel segmentation method based on scale-space analysis of obtaining the vessel geometrical features by the first and the second derivative of the intensity in the image. Then they used a multiple pass region growing procedure which progressively segments the blood vessels. Nain et al.~\cite{nain2004vessel} combined image statistics and shape information to derive a region-based active contour that segments tubular structures and penalizes leakages. Liao et al.~\cite{liao2013globally} introduced a fast marching approach with curvature regularization for vessel segmentation, since most vessels have a smooth path and curvature can be used to distinguish desired vessels. Florin et al.~\cite{florin2006globally} proposed a particle filter based propagation approach for the segmentation of vascular structures in 3D volumes. To obtain posterior probability estimation of the vessel location, Wang et al.~\cite{wang2013sequential} employed sequential Monte Carlo tracking and proposed a vessel segmentation method by fusing multiple cues extracted from CT images for enhanced segments from global path minimization. Forkert et al.~\cite{forkert20133d} presented and evaluated a level-set segmentation approach with vesselness-dependent anisotropic energy weights, which focuses on the exact segmentation of malformed as well as small vessels from time-of-flight (TOF) MRA datasets. Ye et al.~\cite{ye2013noncontrast} proposed non-linear subtraction (NLS) method~\cite{ye2013noncontrast}, which is employed for selective MRA enhancement utilizing the flow rephrased and dephased images. Then the vessel label can be obtained based on an enhanced angiography map. Govyadinov et al.~\cite{govyadinov2018robust} described a template-based predictor-corrector method for tracing filaments that is robust in microvascular datasets, and applied a number of glyph-based visualization techniques to represent the aggregated and biologically relevant information of the extracted microvascular network. Then, they developed a bi-modal visualization framework~\cite{govyadinov2019graph}, leveraging graph-based and geometry-based techniques to achieve interactive visualization of microvascular networks. However, these approaches are exhausted by handcrafted features (e.g., gradients of the intensity, second order local structures, maximum principal curvatures) and complicated manual parameter adjustment to adapt to the subject variations. Therefore, their robustness and accuracy across subjects are limited.
\vspace{-1mm}
\subsection{Data-Driven Vessel Extraction and Segmentation}\vspace{0mm}
Recently, there is an emerging trend to automatically extract, segment, and reconstruct shape objects of interest from input 2D / 3D images~\cite{fan2017point,wang2018pixel2mesh,wang2019deeporgannet} or 3D meshes / point clouds~\cite{masci2015geodesic,qi2017pointnet,xu2017directionally,jin2018learning,komarichev2019cnn} by deep neural network (DNN)~\cite{bronstein2017geometric,book:Hua_spectral2019}. Particularly for vessel structures, several deep learning based methods have been proposed to extract vessels from 2D retinal images. DeepVessel~\cite{fu2016deepvessel} addresses retinal vessel segmentation as a boundary detection task that is solved using a CNN with a side-output layer to learn discriminative representations, and a conditional random field (CRF) layer that accounts for non-local pixel correlations. Li et al.~\cite{li2015cross} presented a supervised method for vessel segmentation by using the cross-modality data transformation from retinal image to vessel map. Mo and Zhang~\cite{mo2017multi} developed a deep supervised fully convolutional network by leveraging multi-level hierarchical features of the deep networks for retinal vessel segmentation. Liskowski and Krawiec~\cite{liskowski2016segmenting} proposed a supervised segmentation technique that uses a DNN trained on a large number of samples preprocessed with global contrast normalization, zero-phase whitening, and augmented using geometric transformations and gamma corrections. Shin et al.~\cite{shin2019deep} incorporated a graph neural network (GNN) into a unified CNN architecture to jointly exploit both local appearances and global vessel structures. Their framework has been evaluated on retinal image datasets and a coronary artery X-ray angiography dataset. These methods can perform well on the 2D vessel segmentation task, but are far from satisfaction / feasibility on 3D micro vessel scenario, since their designs either do not consider the correlation / inter-information between slices in 3D volumetric images or cannot afford the computational and memory burdens in the large 3D volume at the micro-level.

As for deep learning-based 3D vessel segmentation, for instance, Uception~\cite{sanchesa2019cerebrovascular} presents a network inspired by the 3D U-Net~\cite{cciccek20163d} and the Inception modules~\cite{szegedy2017inception} for segmentation of the cerebrovascular network in MRA images. DeepVesselNet~\cite{tetteh2018deepvesselnet} and VesselNet~\cite{kitrungrotsakul2019vesselnet} propose 2D orthogonal cross-hair filters in all sagittal, coronal, and axial planes on each voxel to make use of 3D context information at a reduced computational burden and memory cost. However, the challenging problems in 3D micro-cerebrovascular segmentation are complicated vessel geometry and topology, high sparseness and noise of vessel data in a large-sized 3D volume, and the limited resource of 3D microvascular datasets. The above methods do not overcome these challenges.
\vspace{-1mm}
\section{VC-Net}
\begin{figure}[t] 
	\begin{center}
		\includegraphics[width=0.9\linewidth]{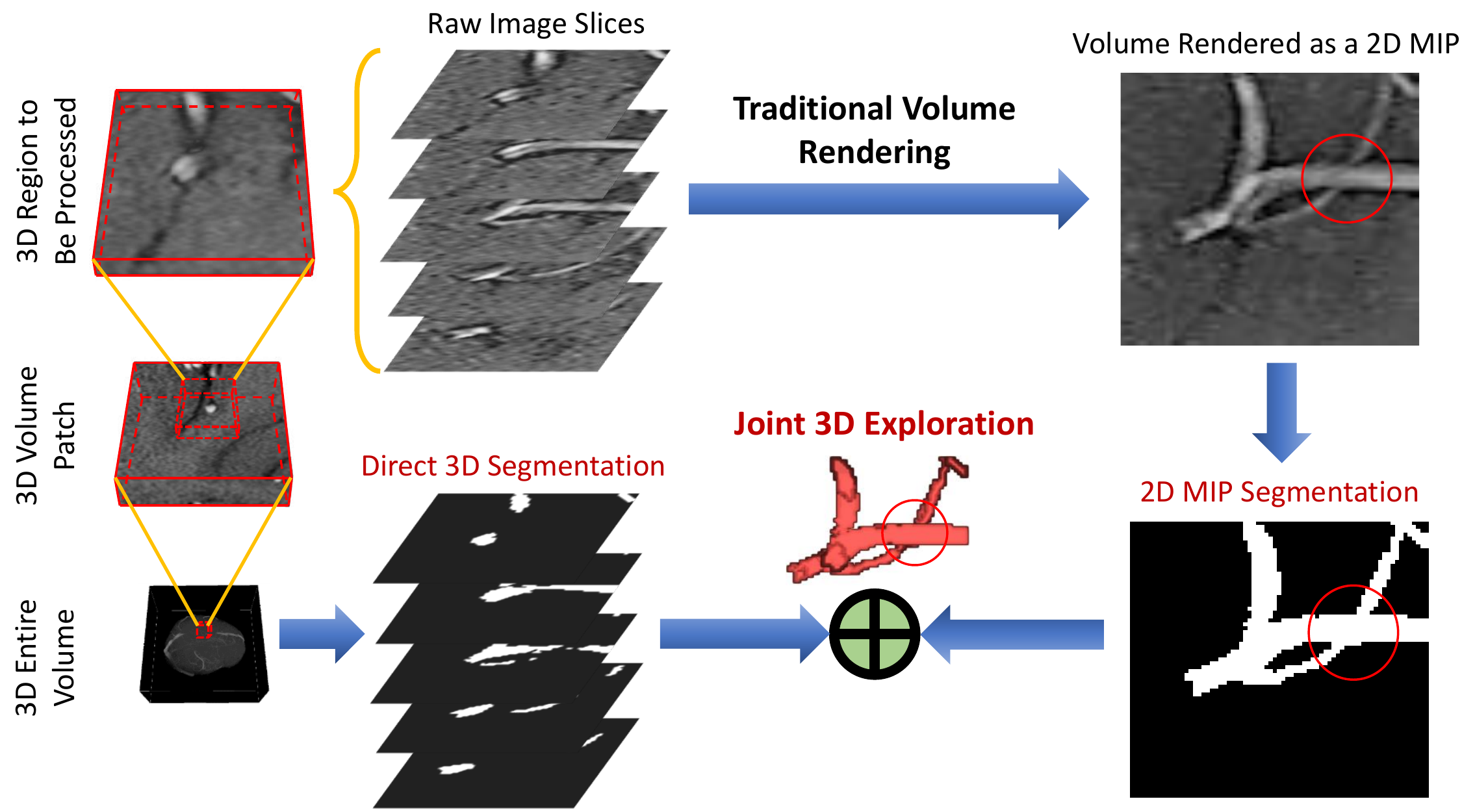}\vspace{-1mm}
		\caption{A novel 3D data analytic paradigm in a joint volume-composition space, where volume rendered results are used to support the visualization-guided 3D volume processing by deep learning.}\vspace{-8mm}
		\label{main_motivation}
	\end{center}
\end{figure}

The fundamental inspiration of the proposed work is to mimic the observation of human exploration in 3D aided by volume rendering. Our work presents a new paradigm to combine direct 3D volume processing and volume rendered clues for effective 3D exploration. For instance in Fig.~\ref{main_motivation}, for a micro-cerebrovascular dataset, the 3D volume image can more accurately represent the 3D spatial information, but the desired task is easy to be confused by challenging SNR and sparse vesselness, as shown in the raw image slices; while the volume rendered 2D MIP image can better enhance the local vessel signal by enforcing vessel continuity and adapt to the geometric variability and scalability of vessels. However, it always lacks 3D spatial sense, e.g., two ``crossing'' and ``kissing'' vessels as circled in red (in fact, the bigger one is above the smaller one in 3D space). So, it is deficient to investigate the (sparse and noisy) 3D data from either 3D volume or volume rendered 2D MIP, respectively. In this work, we design a novel paradigm to support the 3D data analytics, such as segmentation, etc., by using the visualization-guided computing. Instead of conducting the rendering / composition at the final stage as in the traditional visualization pipelines, this paradigm qualitatively investigates the 3D volume data from 2D composited (rendered) images. Essentially, this procedure makes the visualization more important via an early and simultaneous involvement of volume rendering (composition). Finally, we explore the 3D data analytics in a joint volume-composition space. In the following, we introduce the components of the VC-Net model: network architecture and loss function, and dataset generation and preparation.
\vspace{-1mm}
\subsection{Network Architecture}
\label{sec:architecture}
\begin{figure*}[t]
	\begin{center}
		\includegraphics[width=0.9\textwidth]{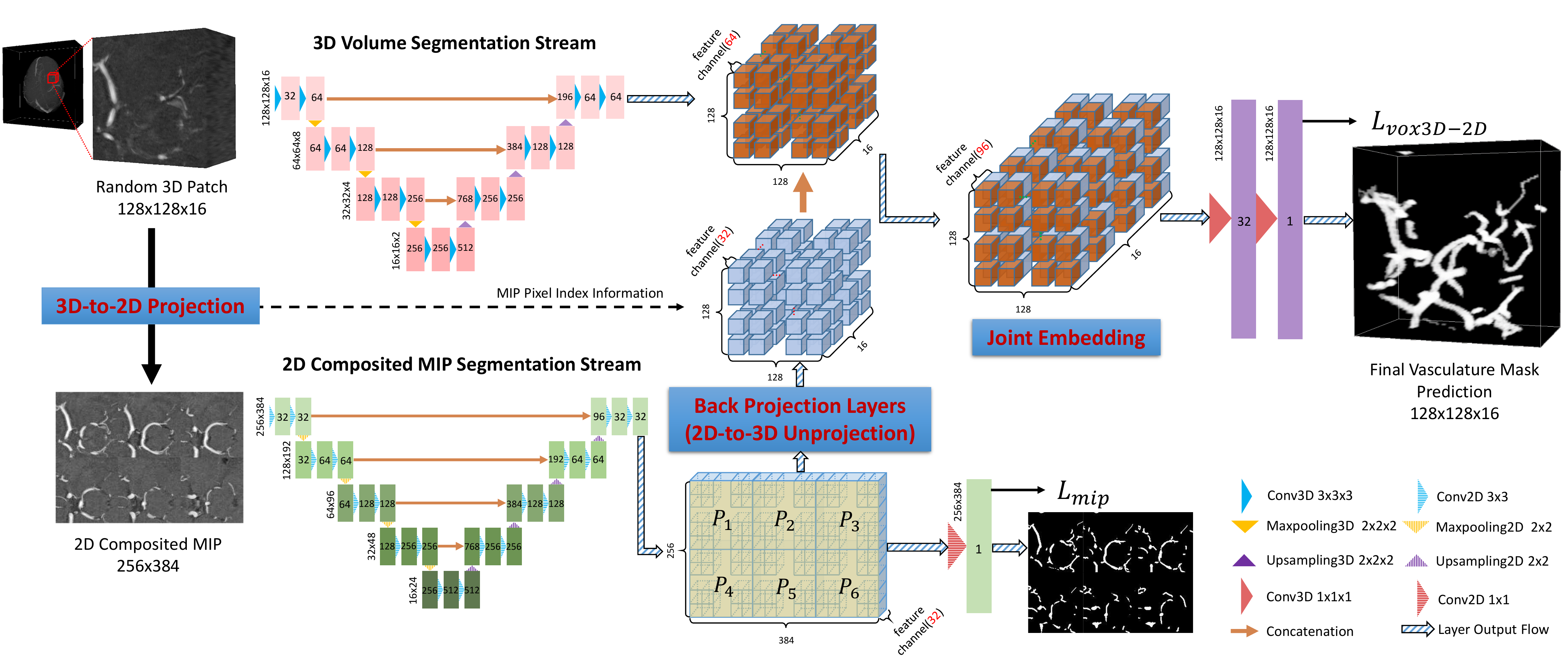}\vspace{-1mm}
		\caption{The architecture of VC-Net. The major procedure includes obtaining the composited MIPs via 3D-to-2D projection, dual-stream segmentation learning for 3D volume and 2D composited MIP feature vectors, back projecting 2D composited MIP feature vectors into the 3D volume feature space via 2D-to-3D unprojection, building a joint convolutional embedding for learning the final vasculature mask.}\vspace{-8mm}
		\label{main_pipeline}
	\end{center}
\end{figure*}

The proposed VC-Net mainly consists of a dual-stream component (i.e., a 3D volume segmentation stream and a 2D composited MIP segmentation stream) and the bi-directional operations between these two streams (i.e., 3D-to-2D projection and 2D-to-3D unprojection). The overall architecture is demonstrated in Fig.~\ref{main_pipeline}. The two-stream segmentation component can learn vessel feature vectors in 3D volume and corresponding multiple 2D MIPs (enhanced and dense depiction of 3D relationships via a 3D-to-2D projection computation) contexts, respectively. After that, the embedded features from the 2D composited MIP are transformed from the 2D MIP domain into the 3D volume domain through a 2D-to-3D unprojection process. Then, the extracted 2D and 3D embedded features from two streams are integrated together, constructing a unified high-dimensional joint convolutional embedding space, which can strengthen the original sparse vessel features from the 3D volume. Finally, the vessel segmentation prediction can be learned at the fusion stage in this joint convolutional embedding space.

In this work, we use a 3D U-Net~\cite{cciccek20163d} as the 3D volume segmentation stream and a half 2D U-Net~\cite{ronneberger2015u} (in terms of feature channel numbers) as the 2D composited MIP segmentation stream, respectively. U-Net-like networks are the most commonly-used and robust medical imaging segmentation neural networks across different data modalities for varying organ / tissue geometries, and thus it is suitable for us to justify the benefits from our 2D-to-3D unprojection and joint embedding of 3D volume and 2D composited MIP. A U-Net-like network is essentially a convolutional encoder-decoder network, which first embeds the input into a high-dimensional feature vector through hierarchical convolution and pooling at the encoder stages, and then decodes the feature vector in the hidden space through hierarchical upsampling and convolution at the decoder stages with the integration of the features directed from different encoder stages through the long-skipped feature concatenations. In Fig.~\ref{main_pipeline}, the layer output feature channel numbers are denoted in the corresponding blocks and layer input spatial dimensions are shown in the horizontal levels of every block.

Due to the limited data availability and large volume size in micro-cerebrovascular image datasets, we choose to train the network patch-wisely. Specifically, from the observation that most brain MRAs have much higher resolutions in axial plane than other planes, we adaptively train our network using non-cubic patches, which have larger dimension size across axial plane, instead of resizing the data into a uniform voxel spacing through an interpolation before the network training, to avoid potential data corruption. As shown in Fig.~\ref{main_pipeline}, the key step in our network is the effective integration of the features from two different streams / domains. In order to fuse the 2D composited MIP stream into the 3D segmentation main stream within the network during the learning process, one may first find out that segmentation task is essentially a dense voxel (pixel)-wise classification problem, and the integration of embedded feature vectors from different learning domains and objectives (i.e., 3D volume segmentation and 2D composited MIP segmentation) must be fused voxel-wisely with the correct spatial correspondence in the volumetric domain. Accordingly, there are two main challenges that need to be overcome in this work. The first one is the effective format of the corresponding 2D composited MIPs from a randomly-extracted 3D volume patch that is adequate and suitable for delivering dense voxel correspondence in the simultaneous dual-stream learning design. The second one is the effective approach for mapping / unprojecting the feature vectors extracted from the composited MIP image plane pixels (in a dimension-reduced 2D) back to the corresponding 3D volume spacing voxels. More details are introduced in following subsections.

\subsubsection{3D-to-2D Projection in Dual-Stream Design}
The major motivation for projecting the 3D volume space into the 2D MIP space is to enhance the local vessel probability. Given a randomly-extracted 3D volume patch $V$ of the size $K_1 \times K_2 \times K_3$ (e.g., we use $128 \times 128 \times 16$ in our experiments) and $K_3$ is along the vertical axis. We compute $s$-sliced (e.g., $s=5$ in our experiments as suggested by domain experts) MIPs of $V$ along vertical axis with overlapping coverage every $t$ slice interval. Consequently we can get a set of $m$ consecutive / sliding MIPs, i.e., $\mathbf{P}=\left \{ P_1, P_2, \ldots, P_k, \ldots, P_{m-1}, P_m \right \}$, in which $P_k$ is the MIP across the $[(k-1)t+1]^{th}$ slice to the $[(k-1)t+s]^{th}$ slice in $V$. It is noted that in a 2D MIP, only one voxel with the maximum intensity among the $s$ voxels along the vertical axis in $V$ will be recorded, which is prone to an information loss, considering the segmentation task actually needs the information of every voxel. Consequently, we set $t=2$ as a trade-off between computation cost and information completeness / denseness. We can get $m$ MIPs of size $K_1 \times K_2$ for $V$, where the MIP number $m$ is computed as:\vspace{-3mm}
\begin{equation}
m = \left \lfloor \frac{1}{t}(K_3-s) \right \rfloor + 1.\vspace{-2mm}
\end{equation}

A MIP conveys denser vessel information and is also naturally suitable for 2D convolution. However, we now have $m$ different MIPs and need to feed them to our network in the MIP stream in company with the 3D volume stream $V$ as an input pair to our entire network. The information from the $m$ MIPs is equally important, which means every pixel information should be kept during learning for later back projection. In order to avoid intuitively stacking them to a $K_1 \times K_2 \times m$ volume such that the 2D CNN (in 2D MIP stream) would essentially treat it as a 2D input of a spatial dimension $K_1 \times K_2$ with $m$ different properties (feature channels), which is essentially deficient in terms of the spatial domain size as well as the operation motivation, we convert the $m$ MIPs to a tiled MIP with a larger 2D spatial size, such as $0.5m K_1 \times 2K_2$. In this case, the 2D convolution is operated equally across the 2D composited MIP plane domain. The slice indices from where the MIP pixels are selected in the original $V$ are also recorded so as to effectively restore the pixel-wise information extracted from MIP to the 3D volume space, which will be used in the 2D-to-3D unprojection in the following process. The format of the 2D composited MIP (e.g., $m=6$ consecutive MIPs) computed from a 3D volume patch is shown in Fig.~\ref{step} (a).

\subsubsection{2D-to-3D Unprojection for Joint Embedding}
Once the 3D volume and 2D MIP streams learn their segmentation features respectively, we intend to integrate them in a unified joint hidden feature embedding space to yield the final 3D segmentation prediction. In order to achieve this, we conduct several operations within our network to unproject (i.e., back project) the pixel features extracted from the composited MIP back to their corresponding 3D voxel feature space.

\begin{figure*}[t]
\begin{center}
\includegraphics[width=0.9\textwidth]{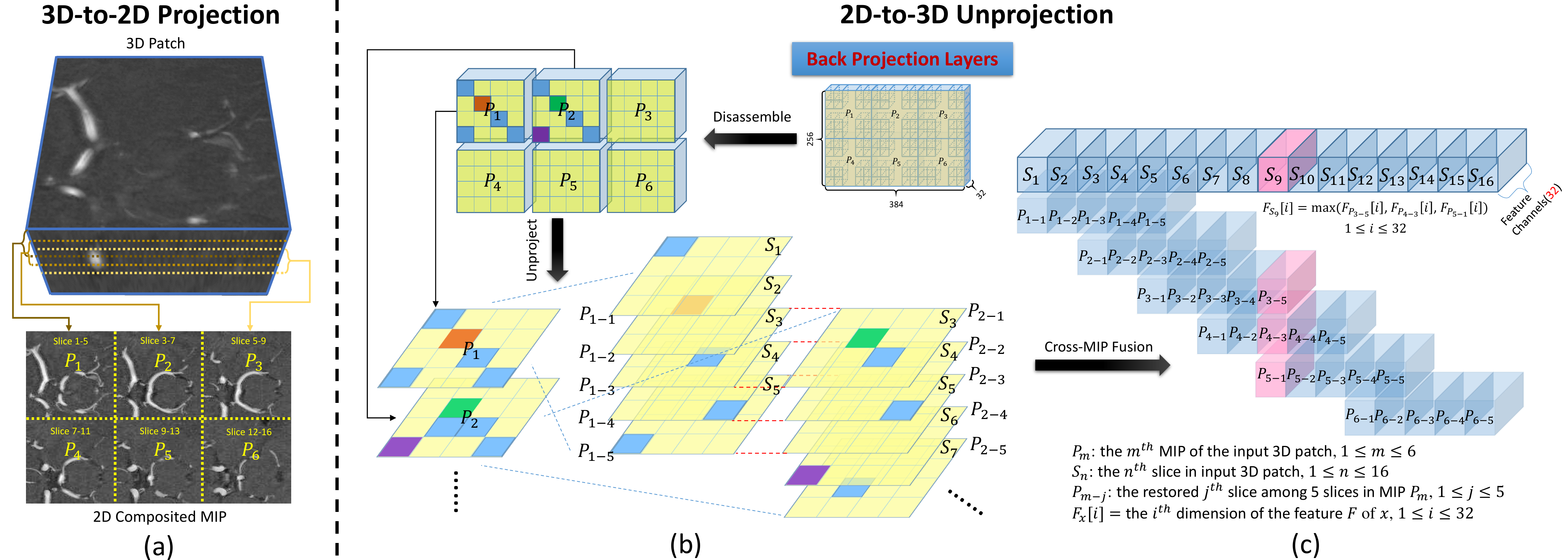}\vspace{-1mm}
\caption{(a) Illustration of the 3D-to-2D projection in the spatial domain for computing a 2D composited MIP from a 3D volume patch. (b) and (c) Illustration of the detailed computations in back projection layers for 2D-to-3D unprojection process in the embedded feature domain.  As illustrated in bottom (b), the consecutive MIPs $P_1$ and $P_2$ with overlapping slice coverage of $S_3$, $S_4$, $S_5$ contribute to information completeness in 3D patch volume. A pixel location on the 5-sliced MIP 2D plane which keeps the feature information of only one voxel out of five (e.g., the middle orange pixel and the left bottom blue pixel in $P_1$ are back projected to $S_2$ and $S_5$) now can be supplemented by $P_2$'s back projection (e.g., the middel green voxel on $S_3$ and the left bottom purple voxel on $S_6$).}\vspace{-8mm}
\label{step}
\end{center}
\end{figure*}

The final-stage hidden feature from 2D composited MIP segmentation stream has the size $0.5m K_1 \times 2K_2$ with $C_1$ channels ($C_1=32$ as shown in Fig.~\ref{main_pipeline}), which is the input of the back projection layers. We first disassemble it to restore $m$ $C_1$-channel features for the corresponding MIPs (e.g., $P_1, P_2, \ldots, P_{m-1}, P_m$, where $m = 6$ as illustrated in Fig.~\ref{step} b). Then we use the recorded index information to map the MIP pixel features back to where they are selected from $V$ during the 2D composited MIP generation. Fig.~\ref{step} (b) shows how the feature vectors of two consecutive MIPs (e.g., $P_1$ and $P_2$) are disassembled from the composited MIP. They unproject their pixel feature space ($P_{m-j}$, i.e., the $j$-th slice among 5-sliced MIP $P_m$, $1 \le j \le 5$) back to the voxel feature space ($S_{n}$, i.e., the $n$-th slice in the input 3D patch, $1 \le n \le 16$). It is noted that the feature dimension is folded from 3D to 2D for a convenient illustration in Fig.~\ref{step} (b) (i.e., hiding the feature dimension).

For the features of overlapping slices (from the consecutive MIPs), which are covered by multiple MIPs, we take the element-wise maximum value across the overlapping restoration through the feature channels:\vspace{-1mm}
\begin{equation}
F_{S_{n}}[i] = \max(F_{P_{1-1}}[i], \ldots, F_{P_{6-5}}[i]), 1 \le i \le 32,\vspace{-0.5mm}
\end{equation}
where $F_{S_{n}}[i]$ represents the $i$-th channel in feature $F$ at the $n$-th slice in the 3D patch. For example, the feature $F_{S_9}$ is computed across the overlapping slices of $P_{3-5}, P_{4-3}, P_{5-1}$ as highlighted in pink in Fig.~\ref{step} (c). The whole process of the cross-MIP fusion in the feature channels of the 3D volume feature space is shown in Fig.~\ref{step} (c) in detail. After that, the unprojected 2D MIP features and 3D volume features from two streams are integrated together, constructing a unified high-dimensional joint convolutional embedding for predicting the final vessel segmentation.

\vspace{-1mm}
\subsubsection{Loss Function}
The major learning objective of our VC-Net is to extract the sparse 3D vasculature structure from the 3D MRI volume image using a 3D segmentation network supplemented by information from multiple denser and more connected 2D MIPs. Consequently the network loss function consists of two terms:\vspace{-1.5mm}
\begin{equation}
L = L_{{vox}_{3D-2D}} + \lambda L_{mip},\vspace{-1mm}
\end{equation}
where $L_{{vox}_{3D-2D}}$ is a joint 3D-2D segmentation Dice loss adopted in 3D volume stream and defined as:\vspace{-1mm}
\begin{equation}
L_{{vox}_{3D-2D}} = - \frac{2\Sigma_{x \in V} p(x) g(x)+\delta}{\Sigma_{x \in V} p(x)+\Sigma_{x \in V} g(x)+\delta},\vspace{-1mm}
\end{equation}
where $p(x)$ and $g(x)$ are the predicted voxel-wise vessel probability maps and ground truth binary labels within the query volume patch $V$, respectively. $\delta$ is a small smooth constant. $L_{mip}$ is applied in 2D composited MIP stream and acts as a regularization term during the learning process, which is also a Dice loss function defined (similarly to $L_{{vox}_{3D-2D}}$) within the 2D composited MIP plane and supervised by the ground truth MIP vessel binary labels. $\lambda$ is the constant coefficient of $L_{mip}$, which is set to be $0.2$ for our best experiment performance.

\subsection{Dataset Generation and Preparation}
\label{sec:datasets}
In this work, we use two different real patient datasets to evaluate our proposed VC-Net method.

\textbf{Novel MICRO-MRI Imaging and Dataset.} Some researchers have recently developed a next generation of microvascular imaging, i.e., Microvascular In-vivo Contrast Revealed Origins Magnetic Resonance Imaging (MICRO-MRI)~\cite{shen2020detecting,Wang2020capability}. Thanks to MICRO-MRI, we became the first ever to be able to acquire such brain imaging datasets and observe the complicated micro cerebral vessels. This dataset is produced by neurologists and radiologists within our collaborative group. Data was acquired with an adapted 3D gradient echo susceptibility weighted imaging (SWI) sequence~\cite{chen2018interleaved} collected from a 3T MR scanner. The post-contrast data were acquired during a gradual increase in dose (final concentration = 4 mg / kg). Eleven healthy volunteers were scanned in brain regions with a dual echo SWI sequence at four time points: the first was acquired pre-contrast and the remaining three were acquired post-contrast during a gradual increase in dose delivered over the time frame of 20 min; with the imaging parameters: echo time (TE)1 / TE2 / repetition time (TR) = 7.5 / 22.5 / 27 ms, bandwidth = 180 Hz / pxl, flip angle = $15^{\circ}$ (pre-contrast and final post-contrast data) and $20 ^{\circ}$ (first and second post-contrast data). The voxel spacing is $0.22 \times 0.22 \times 1$ $\text{mm}^3$ with a volume size of $1024 \times 832 \times 96$ voxels.

\emph{Major-level vessel data.} This protocol enables multiple image sources for producing both MR arteriogram (MRAG) and venogram (MRVG). For the MRVG, the pre-contrast quantitative susceptibility mapping (QSM) and ${\text{R}_{\text{2}}}^*$ constitute two different representations of veins. In order to obtain the pre-contrast QSM data, the original phase data was unwrapped using the 3D best path method~\cite{abdul2007fast}. The sophisticated harmonic artifact reduction for phase data (SHARP) method was used to estimate the background field and remove it from the unwrapped phase~\cite{schweser2011quantitative}. The truncated k-space inverse filter approach with an iterative geometric constraint (also known as iSWIM) was applied to the resultant phase to generate the QSM data~\cite{haacke2010susceptibility,tang2013improving}. The QSM data was further refined by removing the strong phase gradients from the long TE phase data based on a quality phase mask. The resultant phase was used to obtain a QSM data $\text{QSM}_{\text{TE2}}$. The QSM of the short TE data $\text{QSM}_{\text{TE1}}$ was also generated, but without using a quality map, since at a low TE the phase gradients were not that strong. Finally, the missing information on $\text{QSM}_{\text{TE2}}$ was filled in by applying an inverted quality mask to $\text{QSM}_{\text{TE1}}$. To obtain the pre-contrast ${\text{R}_{\text{2}}}^*$, the short and long TE magnitude data $\mathbf{S}(t)$ were fitted to the monoexponential equation: $\mathbf{S}(t) = \rho e^{-(t {\mathbf{R}_2}^*)}$, where $\rho$ is the tissue intrinsic proton density.

Another MRVG was generated by subtracting the short TE magnitude data of pre-contrast from the short TE magnitude data of the first post-contrast. The above-mentioned subtraction provides a venous-only map $\text{V}_{\text{T}_\text{1}}$. The $\text{QSM}$, ${\text{R}_{\text{2}}}^*$, and $\text{V}_{\text{T}_\text{1}}$ maps were then normalized to values between 0 and 1, and an average of these different sources produced a high-quality MRVG referred to as $\mathbf{MRVG}_{\text{avg}}$.

An MRAG was then calculated using a nonlinear subtraction (NLS)~\cite{ye2013noncontrast}, i.e., $\mathbf{MRAG}_{\text{nls}}$, of the long TE $\mathbf{S}'$ from the short TE $\mathbf{S}$ of the pre-contrast magnitude data as: $\mathbf{MRAG}_{\text{nls}} = \mathbf{S}^2 - \alpha \mathbf{S}'^2$, where $\alpha$ is a constant with an empirically selected value of $1.5$. Due to the ${\text{T}_{\text{2}}}^*$ effect, this subtraction also enhances the veins, but to a much smaller extent than the arteries. Nevertheless, any venous enhancement is discarded by using a mask generated from $\mathbf{MRVG}_{\text{avg}}$. Finally, the ultimate ground truth vessel labels are obtained by integrating the enhanced angiography (i.e., arteriogram and venogram) maps~\cite{buch2020subvoxel} from the computed $\mathbf{MRAG}_{\text{nls}}$ and $\mathbf{MRVG}_{\text{avg}}$, with a threshold-based method for the initial masks, followed by domain experts' post-manual labeling refinement using our developed cerebrovascular labeling and visualization tool. Supplemental Material and Video are included for demonstrating the interactive interface and basic functions in detail.

\emph{Micro-level vessel data.} SWI images were generated by homodyne high-pass filtering (filter size = $96 \times 96$) the phase images to generate a phase mask, which was multiplied with the original magnitude images four times, for all time points~\cite{haacke2004susceptibility}. All the original magnitude and corrected phase data were then registered to the pre-contrast data. The short TE (7.5 ms) magnitude data of pre-contrast and the first post-contrast time points were averaged. This averaged magnitude data was subtracted by the long TE (22.5 ms) SWI data from the last post-contrast time point (4 mg / kg) to enhance the vessels. The vessels were further enhanced on the resultant subtracted image by applying the vesselness algorithm~\cite{frangi1998multiscale} to obtain the micro-level vessel map. The micro-level vessels from this resultant vessel map were extracted using an adaptive threshold-based region growing method (ATRG)~\cite{jiang2005argdyp}, i.e., $\mathbf{SWI}_{\text{ATRG}}$, as the initial masks, followed by domain experts' manual inspection of the extracted vessels for quality control.

\textbf{Public MRA Dataset.} In order to compare with the existing methods, we use a public TubeTK Toolkit MRA dataset from University of North Carolina at Chapel Hill~\cite{TubeTK}, acquired by a 3T MR system. There are 42 patient cases in the whole dataset, which have the manual-labeled vessel segmentation masks. The voxel spacing of the MRA images is $0.5 \times 0.5 \times 0.8$ $\text{mm}^3$ with a volume size of $448\times448\times128$ voxels.

\vspace{-2mm}
\section{Results}
For both MRA TubeTK and MICRO-MRI datasets (the different modalities of input image examples are provided in Supplemental Material), we first apply the MR-based skull-stripping method~\cite{Jenkinson} to extract the pure brain from each image. As we mentioned in Sec.~\ref{sec:architecture}, our VC-Net network is designed for patch-wise training and the 3D training patches with the imbalanced dimensions are randomly-extracted with overlapping focusing on the brain area in the whole 3D MRA / MICRO-MRI, e.g., 80 patches for each TubeTK case and 440 patches for each MICRO-MRI major-level vessel case. The random training / validation / testing case split is 33 / 3 / 6 and 6 / 2 / 3 for the TubeTK dataset and the MICRO-MRI major-level vessel case, respectively. All the numerical evaluations are reported in terms of whole brain volume image patched with no overlapping.

Our VC-Net adopts the Adam optimizer with 0.0001 as an initial learning rate, 0.5 as the learning decay factor, and 10 epochs as the learning patience across all datasets. In our implementation, we restrict our batch size to 4 due to hardware limits. No batch normalization is adopted in either stream in VC-Net and we use ReLU (Rectified Linear Unit) activation for both 2D and 3D convolutional layers in corresponding streams and sigmoid activation for the final vessel probability output from both 2D and 3D streams. The network is implemented in TensorFlow framework and the total training time is around 10 hours on two NVIDIA GeForce GTX 1080 GPUs with 8 GB GDDR5X memory. The inference time is given in the following subsection. Data and source code of this work will be made available.

The performance of our VC-Net and all methods in comparison are numerically evaluated by the following three quantitative metrics, which are defined from the classifier confusion matrix from different aspects:

\textbf{Dice Similarity (Dice)}, $2TP/(2TP+FP+FN)$, (the same as F-score under most of the circumstances) generally measures the intersection over union between prediction and ground truth. It involves true positive ($TP$), false positive ($FP$), and false negative ($FN$), so as to be the most comprehensive indicator to evaluate the sparse vessel segmentation in a large portion of background, i.e., true negative ($TN$).

\textbf{Precision}, $TP/(TP+FP)$, measures the model ability of ruling out the noise contributions and obtaining the correct vessel voxels.


\textbf{False Positive Rate (FPR)}, $FP/(FP+TN)$, examines the model ability of distinguishing the real background and noise against vessels, which is crucial for the clinical purpose.

%
%
%

The best results in tables are shown in bold font. Here we do not include the metric of Accuracy, due to its extremely high value (e.g., $\ge 99\%$) for all methods. The reason is that it involves dominant portion of background (true negative) together with highly sparse target (e.g., the segmented vessels in our task) in computation and consequently loses its effectiveness for segmentation evaluation.
\vspace{-1mm}
\subsection{Comparison with the State-of-the-Art}
\label{sec:comparison}
We first compare our VC-Net performance on TubeTK dataset with four state-of-the-art deep learning based methods (i.e., 3D U-Net~\cite{cciccek20163d}, 2D U-Net~\cite{ronneberger2015u}, DeepVesselNet~\cite{tetteh2018deepvesselnet}, and Uception~\cite{sanchesa2019cerebrovascular}) and one classical parametric intensity-based method (i.e., vesselness algorithm~\cite{frangi1998multiscale,descoteaux2008geometric}) in 3D vessel segmentation. All deep learning methods in comparison are trained until convergence by using the same dataset split or using the results reported from their original publication (such as Uception). For 2D U-Net, we train it with $128 \times 128$ 2D patches, whose amount is over 10 times of the 3D patch amount extracted for the 3D CNN based methods in comparison with on-the-fly data augmentation for a fair data acquisition. For DeepVesselNet, we have tried different combinations of their data pre-processing process and chosen the image intensity clipping for obtaining an optimal performance on TubeTK dataset.

\begin{figure*}[t]
	\begin{center}
		\includegraphics[width=0.85\textwidth]{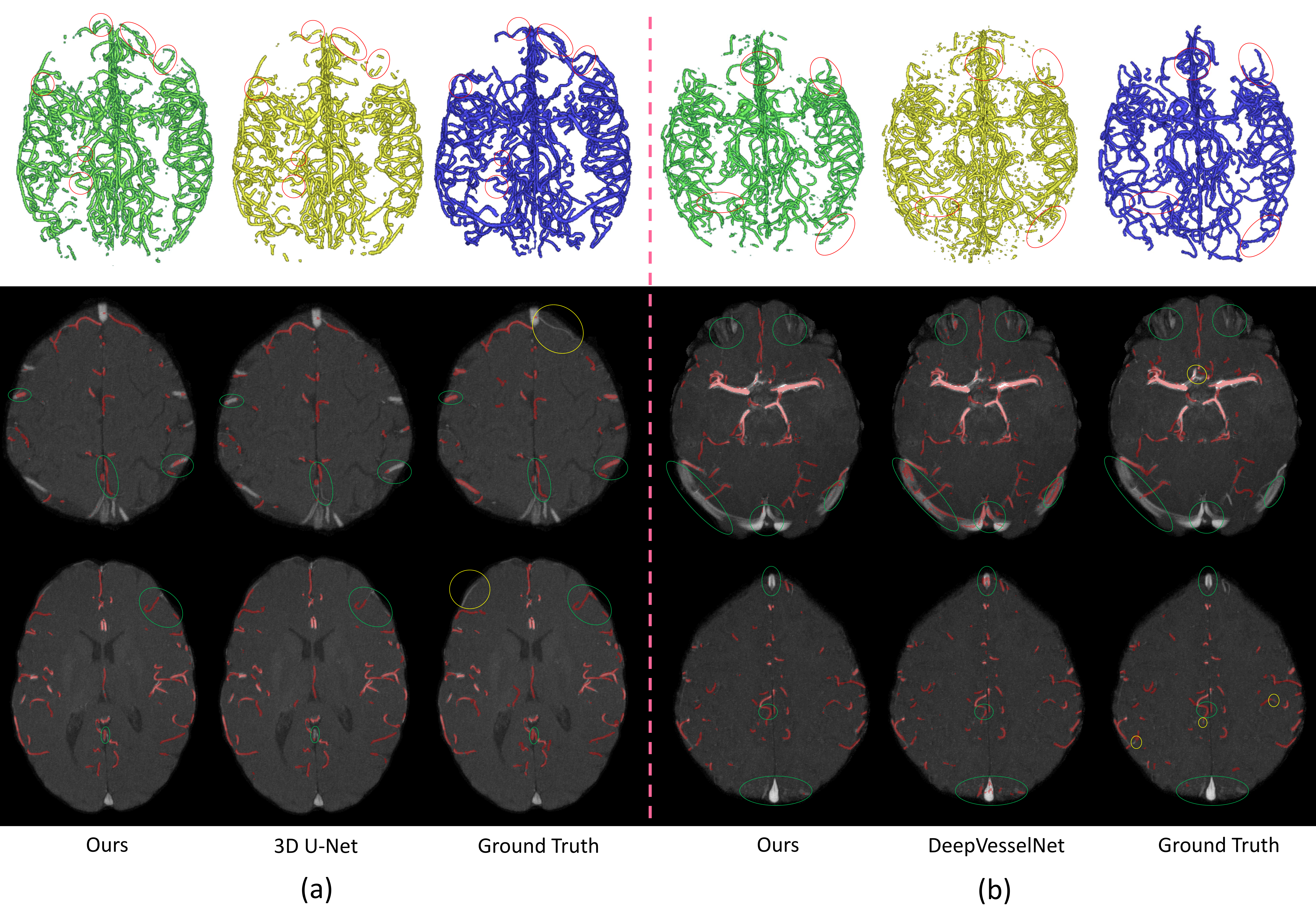}\vspace{-2.5mm}
		\caption{Some qualitative comparison results from TubeTK dataset: The 3D global vessel segmentations are shown from superior direction. The MIP segmentations are visualized by 5-sliced MRA images, and the corresponding vessel masks in MIPs are marked in semi-transparent red. The highlighted comparison areas are marked in circles. Yellow-circled areas are some minor mistakes in the ground truth (discussed in Sec.~\ref{sec:discussion}).}\vspace{-6mm}
		\label{visual1}
	\end{center}
\end{figure*}

The quantitative performance comparison of these methods on TubeTK dataset is shown in Tab.~\ref{tab1}. `$-$' means `not applicable' due to lack of their implementations or results. Here we also provide the per-volume inference time and the parameter number to evaluate the model efficiency besides the segmentation performance. From Tab.~\ref{tab1}, we can see that our VC-Net has overall the best segmentation performance among all the methods on TubeTK dataset. With the 2D composited MIP feature integration, our network performs better than a pure 3D U-Net~\cite{cciccek20163d} over the three different metrics on segmentation results. The qualitative comparison of MIP-wise (e.g., 5-sliced) segmentation results and 3D global vessel segmentation results between our VC-Net and 3D U-Net (one of the most robust state-of-the-art deep learning based methods for biomedical image segmentation) is shown in Fig.~\ref{visual1} (a). With the 2D composited MIP complementary information, the final vessel segmentation shows better connectivity and better small vessel capturing as marked in red circles (3D global vessel segmentation visualization) and green circles (2D MIP vessel segmentation visualization). Besides the segmentation performance gain, the increase of time and space complexities in VC-Net is not high compared with a standalone 3D U-Net as shown in Tab.~\ref{tab1}, since only a half 2D U-Net (i.e., 7.8 million parameters) is involved in the 2D MIP stream. On the other hand, since the computational complexity of 3D convolution operations apparently overweighs that of 2D convolution operations, the 3D stream still dominates the computational complexity of the entire VC-Net. Another observation is that the 3D U-Net greatly outperforms 2D U-Net~\cite{ronneberger2015u} even if the latter contains many more feature embedding channels, since the former method is able to capture the cross-slice continuity and that is why 3D CNN should be involved in such sparse 3D object segmentation with complex topology. Moreover, the full 2D U-Net implies much larger amount of 2D convolution operations and model parameters, which lead to the unsatisfactory model efficiency. DeepVesselNet~\cite{tetteh2018deepvesselnet} fails to yield a good performance as they reported in their own dataset, which could result from the lack of the pre-training procedure, i.e., a relatively complicated data pre-processing, and the instability of their loss function, which is severely sensitive to the training perturbation. In addition, their light-weighted network only consists of five convolutional layers for high efficiency, whereas its simplicity may undermine its cross-dataset robustness. The qualitative comparison with DeepVesselNet is given in Fig.~\ref{visual1} (b), from which you can see that the DeepVesselNet result is much noisier and has severer connectivity issues. Our method also outperforms the best Uception result reported in~\cite{sanchesa2019cerebrovascular} on TubeTK dataset with even less data pre-processing procedure, which implies that empirical neural network modification to increase the model complexity does not guarantee a better performance all the time. To be comprehensive, we apply the vesselness algorithm~\cite{frangi1998multiscale,descoteaux2008geometric} as a traditional benchmark method for comparison based on the available data modality in TubeTK dataset, which is a widely-used approach to segment the cylindrical vessel structures in medical field. As shown in Tab.~\ref{tab1}, all deep learning methods greatly outperform classical vesselness method on TubeTK dataset with much higher efficiency, which is beneficial from the essence of deep learning techniques: mostly end-to-end, more adaptive non-linearity, controllable ambiguity for feature extraction and integration process than traditional complex manual parameter-driven algorithms to boost the overall robustness and accuracy for real-world image learning tasks. More qualitative comparisons with 3D U-Net and DeepVesselNet are provided in Supplemental Material.\vspace{-2mm}


\begin{table}[htb!]
	\caption{Quantitative performance evaluation of different methods on TubeTK dataset.}\vspace{-1mm}
	\label{tab1}
    \centering
	\begin{adjustbox}{width=\columnwidth,center}
		\begin{tabular}{|c|c|c|c|c|c|c|}
			\hline
			Methods / Metrics & Dice (\%) $\uparrow$ & Precision (\%) $\uparrow$ &  FPR (\%) $\downarrow$ &   Time (s) $\downarrow$ & \# Para. $\downarrow$\\
			\hline \hline
			Ours & \textbf{71.81} & \textbf{76.66}  &  \textbf{0.0821} &  10.8 & 24 M \\
			3D U-Net & 71.01 & 74.00 &   0.0958 &  7.5 & 19 M\\
			Uception  & 67.01 & $-$ &  $-$ & $-$ & $-$\\
			DeepVesselNet & 64.12 &63.75 & 0.1465 &  \textbf{1.8} & \textbf{0.06 M}\\
			2D U-Net & 65.10 & 70.05 &  0.1041 &  16.7 & 31 M\\
			Vesselness & 37.71 & 47.69 &  0.1393 &  186.6 & $-$\\
			\hline
		\end{tabular}
	\end{adjustbox}
\end{table}
\vspace{-3mm}

\vspace{-1mm}
\subsection{Experiments and Evaluation on MICRO-MRI Dataset}
\label{sec:micro_mri}
As mentioned in Sec.~\ref{sec:relatedwork}, most of the recently related work on brain vasculature segmentation tasks is limited to major arteries (in Sec.~\ref{sec:comparison}) since currently most of the available brain MRI image datasets with adequate amount and consistent quality are MRAGs. However, with assistance from the neurologists and radiologists under our collaboration, we can now extend VC-Net from general artery segmentation to the vasculature segmentation of major artery and major vein, separately. More inspiringly, we also demonstrate that our VC-Net is capable of extracting the micro vessels in the complicated real patient MICRO-MRIs. It is noted that the segmentation of brain vessels becomes more challenging in micro-level than major-level, and more challenging in veins than arteries. The following experiments show that our method has bigger improvements on more challenging cases (i.e., micro-level vessel and major-level vein segmentations) compared with other methods.

\subsubsection{Major-Level Artery Segmentation and Visualization}
\begin{figure*}[htb!]
	\begin{center}
		\includegraphics[width=0.85\textwidth]{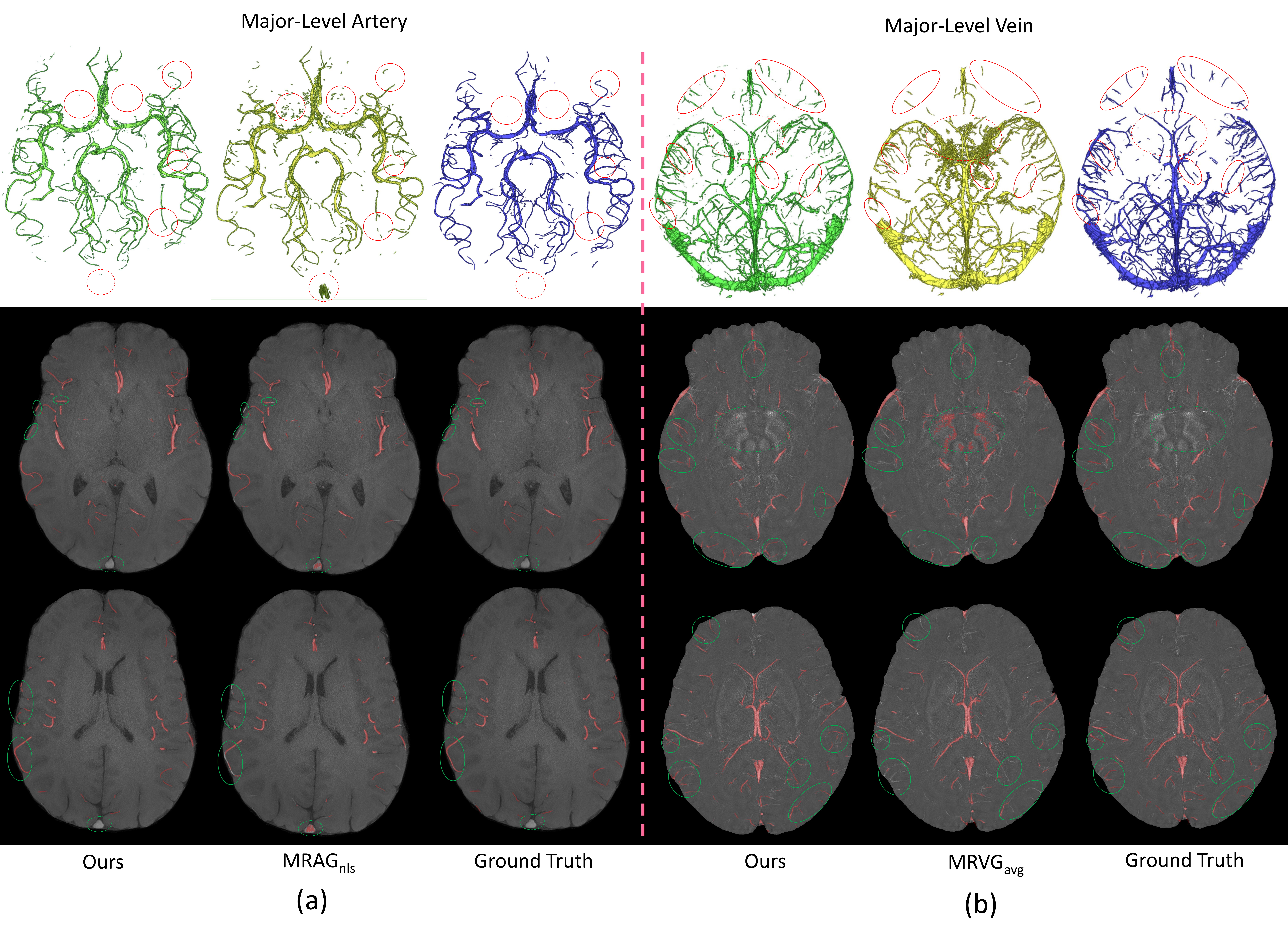}\vspace{-2.5mm}
		\caption{Some qualitative comparison results from MICRO-MRI major-level vessel dataset: The 3D global vessel segmentations are shown from superior direction. The MIP segmentations are visualized by 5-sliced MICRO-MRI images, and the corresponding vessel masks in MIPs are marked in semi-transparent red. The highlighted comparison areas are marked in circles. The 3D MRAG / MRVG images from MICRO-MRI dataset only focus on midbrain area and thus have less vessels compared with TubeTK dataset.}\vspace{-6mm}
		\label{visual2}
	\end{center}
\end{figure*}

The MRAGs in clinic MICRO-MRI datasets under our collaboration as mentioned in Sec.~\ref{sec:datasets} focus on midbrain area from where the major-level vessels are relatively denser and more observable. Currently our collaborative domain experts apply the state-of-the-art model-driven NLS method~\cite{ye2013noncontrast} (i.e., $\text{MRAG}_{\text{nls}}$) followed by case-wise threshold selection to extract the clean midbrain vessels, which requires different data modalities and tedious manual parameter-tuning as stated in Sec.~\ref{sec:datasets}. However, from Fig.~\ref{visual2} (a) we can see that its segmentation result still fails to be free from location-dependent interference, such as superior sagittal sinus (red dotted circles in 3D visualization and green dotted circles in MIP visualization) and some random scattered voxel noise (red solid circles). Aiming to improve the segmentation performance with less manual-parameter tuning and less modality requirement yet provide a much more efficient method for major artery extraction that is well applicable for future patient case collection, we train our VC-Net with only TE1 pre-contrast SWI (a single-modal MRAG) data as input. From Tab.~\ref{tab2} we can see that our quantitative evaluation results outperform the $\text{MRAG}_{\text{nls}}$ method on all metrics even if the latter one integrates and enhances artery signal from several different data modalities. Here we also include our numerical comparison with 3D U-Net (under the same experiment setting), the most competitive method on TubeTK dataset to show our network's cross-dataset robustness and superiority. One can also observe that the numerical difference of the performance in MICRO-MRI major-level artery dataset is not as obvious as the major-level vein and micro-level vessel datasets as shown in the following two subsections, which may result from the fact that the MRAGs are relatively clearer in terms of the image dose effect and the noise type. More qualitative comparisons with the $\text{MRAG}_{\text{nls}}$ method are provided in Supplemental Material.\vspace{-1mm}

\begin{table}[htb!]
	\caption{Quantitative performance evaluation of different methods on major-level artery segmentation.}\vspace{-1mm}
	\label{tab2}
    \centering
	\begin{adjustbox}{width=0.7\columnwidth,center}
		\begin{tabular}{|c|c|c|c|c|}
			\hline
			Methods / Metrics & Dice (\%) $\uparrow$ &  Precision (\%) $\uparrow$ &  FPR (\%) $\downarrow$  \\
			\hline \hline
			Ours & 	\textbf{82.98} & \textbf{83.69} &  \textbf{0.0337}  \\
			3D U-Net & 82.64  &83.52  &  0.0342  \\
			$\text{MRAG}_{\text{nls}}$ & 80.60 & 82.07  &  0.0354 \\
			\hline
		\end{tabular}
	\end{adjustbox}
\end{table}

\vspace{-1mm}
\subsubsection{Major-Level Vein Segmentation and Visualization}
Currently, the MRVGs are not readily and directly available from the scanner, so there is no raw MRI image that can produce pure veins. There are different ways that people have used to derive it such as the SWI, QSM, or ${\text{R}_{\text{2}}}^*$ data, where the veins are highlighted. However, they all have background tissues as well as noise associated with them. In this work, the MRVGs from the MICRO-MRI dataset are ultimately acquired through the $\text{MRVG}_{\text{avg}}$ method by enhancing vein signals from different data resources as mentioned in Sec.~\ref{sec:datasets}. Our collaborative domain experts compute the vein labels by post-manual case-wise threshold adjustment on the MRVGs. However, we can see from the major-level vein case in Fig.~\ref{visual2} (b) that the vein labels still fail to be free from artery artifacts as marked in red / green dotted circles. In addition, the MRVGs overall have more challenging noise type (e.g., very strong artery artifacts) than single-modal MRAGs due to modality formulation; therefore, the corresponding intensity-based vein labels tend to have more fuzzy edges. However, our VC-Net is able to effectively overcome the aforementioned difficulty as shown in Fig.~\ref{visual2} (b). Tab.~\ref{tab3} shows the numerical comparison among ours, $\text{MRVG}_{\text{avg}}$, and 3D U-Net under the same experiment setting. We can see that our numerical results overall outperform the other two methods. The general lower numerical performance compared to artery segmentation (in the previous subsection) may result from more challenging input data type and the sinus region (i.e., the large and thick vein-like area at the bottom in 3D global segmentation). More qualitative comparisons with the $\text{MRVG}_{\text{avg}}$ method are provided in Supplemental Material.\vspace{-1.1mm}

\begin{table}[htb!]
	\caption{Quantitative performance evaluation of different methods on major-level vein segmentation.}\vspace{-1mm}
	\label{tab3}
    \centering
	\begin{adjustbox}{width=0.7\columnwidth,center}
		\begin{tabular}{|c|c|c|c|c|}
			\hline
			Methods / Metrics  &  Dice (\%) $\uparrow$ & Precision (\%) $\uparrow$ &  FPR (\%) $\downarrow$ \\
			\hline \hline
			Ours  & \textbf{76.46} &\textbf{82.20} &  \textbf{0.0849} \\
			3D U-Net & 76.02 &80.40 &  0.0946  \\
			$\text{MRVG}_{\text{avg}}$ & 73.73 &64.99 &  0.2294 \\
			\hline
		\end{tabular}
	\end{adjustbox}
\end{table}

\vspace{-1mm}
\subsubsection{Micro-Level Vessel Segmentation and Visualization}
\begin{figure*}[htb!]
	\begin{center}
		\includegraphics[width=0.85\textwidth]{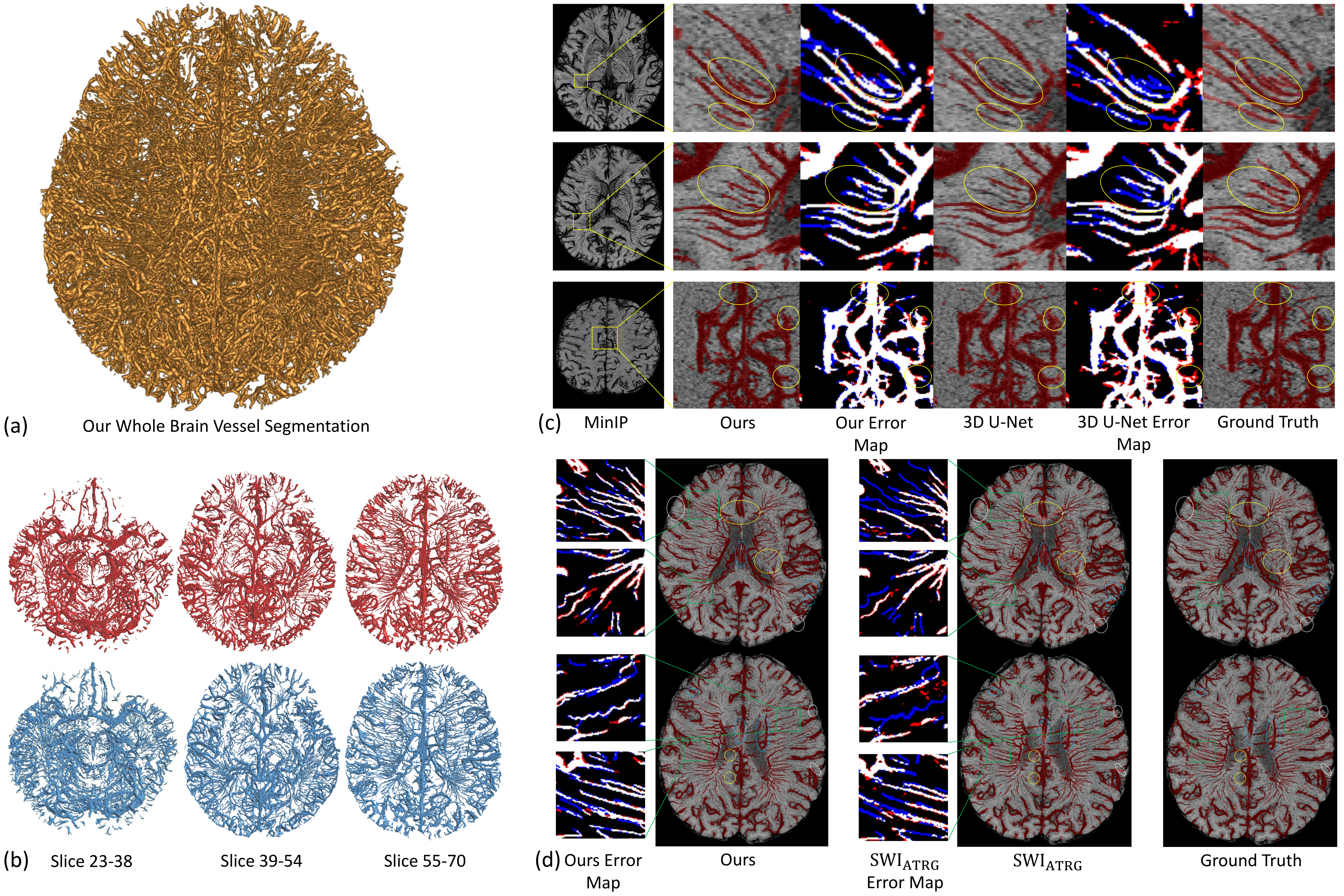}\vspace{-2mm} 
		\caption{(a) Whole brain micro vessel segmentation result of our method from superior direction. (b) Midbrain non-overlapping subarea detail visualization (in red) with ground truth in comparison (in blue). (c) Qualitative comparison results between our method and 3D U-Net on MICRO-MRI dataset, shown as three patch details extracted from three different 5-sliced MinIPs and their corresponding error maps on micro-level and major-level vessels. (d) Qualitative comparison results between our method and the $\text{SWI}_\text{ATRG}$ method on MICRO-MRI dataset, shown as two 5-sliced MinIP segmentations and their corresponding error maps on micro-level vessels. The highlighted comparison areas are marked in circles.}\vspace{-6mm}
    \label{lvl2}
	\end{center}
\end{figure*}

As mentioned in the introduction, the micro-cerebrovasculature turns out to be a good physical indicator of many neurological disorders and vascular diseases; thus it is extremely important and a breakthrough for MICRO MRAV diagnosis of vascular disease to trace small vessels and analyze their topology, morphology, density, and distribution with direct visual inspection of microvascular abnormalities \emph{in-vivo}. Besides the major-level vessel segmentation, our VC-Net shows great capability to track the micro vessels as well. In this experiment, it is noted that we have the input data modality format which is different from those in the previous experiments as shown in the first column in Fig.~\ref{lvl2} (c), i.e., the minimum intensity projection (MinIP) images. The different modalities of input image slice examples in the experiments are provided in Supplemental Material. All vessels, including major and micro ones, appear to be dark (very low voxel intensity) and show no contrast to the dark background, which may cause confusion to our network in the 2D composited MIP segmentation stream even if we accordingly switch to compute the MinIP instead. In order to keep the framework consistency and take advantage of the pre-trained network in the previous subsections, we inverse the voxel intensity within the brain foreground area in the whole 3D SWI image and then extract 1320 random patches from each training image (considering the vessels are much denser in a large-sized 3D volume, more patches per case can make up for limited datasets, e.g., two training cases). By fine-tuning our VC-Net pre-trained on major-level MRI images with these patches, our network is capable of capturing the continuous micro vessels clearly as shown in Fig.~\ref{lvl2}.

Currently, SWI is the only data modality available to capture the micro-level vessels and it includes a large number of major-level vessels as well. Consequently, it is quite challenging to provide a rigorous numerical evaluation on pure micro-level vessels. Alternatively, Tab.~\ref{tab4} shows the numerical evaluation based on the whole SWI image for reference, in which our method quantitatively outperforms 3D U-Net (also fine-tuned from the weights pre-trained on the same major-level MRI images) and the $\text{SWI}_\text{ATRG}$ method on Dice, precision, and FPR metrics. The $\text{SWI}_\text{ATRG}$ method is a state-of-the-art algorithm that our collaborative domain experts are currently using as described in Sec.~\ref{sec:datasets}.

Fig.~\ref{lvl2} (a) and (b) show our whole brain segmentation result (in gold) accompanied by non-overlapping midbrain subarea segmentation results (in red) and their corresponding ground truth (in blue). Fig.~\ref{lvl2} (c) and (d) visualize the qualitative performance of pair-wise comparisons. From Fig.~\ref{lvl2} (d), we can see that the result from the $\text{SWI}_\text{ATRG}$ method suffers severe voxel intensity noise (circled in yellow) and unexpectedly thicker vessels (circled in blue) due to the bold intensity threshold in sacrifice to capture as many micro vessels as possible; however, it still lacks the satisfiable ability of detecting micro vessels as shown in the corresponding zoom-in green patch error maps (white: true positive, red: false positive, blue: false negative, black: true negative). In addition, the $\text{SWI}_\text{ATRG}$ method requires several data modalities which are acquired from different time points as mentioned in Sec.~\ref{sec:datasets}; consequently, the corresponding computed vessel mask also has nonnegligible registration errors (circled in white). However, even if 3D U-Net can alleviate most of the issues that the $\text{SWI}_\text{ATRG}$ method is faced with, it is still insufficient to track the super micro vessel without the 2D MIP complementary information as shown in first two zooming-in patches (circled in yellow) and their corresponding error maps in Fig.~\ref{lvl2} (c). Similar as the $\text{SWI}_\text{ATRG}$ method, 3D U-Net also performs more boldly on covering major-level vessels (less blue on error maps) as shown in the third zooming-in patch in Fig.~\ref{lvl2} (c). However, 3D U-Net incurs more noises (more red on error maps) and thus it has a worse precision.\vspace{-2mm}

\begin{table}[htb!]
	\caption{Quantitative performance evaluation of different methods on micro-level vessel segmentation.}\vspace{-1mm}
	\label{tab4}
    \centering
	\begin{adjustbox}{width=0.7\columnwidth,center}
		\begin{tabular}{|c|c|c|c|c|}
			\hline
			Methods / Metrics &  Dice (\%) $\uparrow$  & Precision (\%) $\uparrow$ & FPR (\%) $\downarrow$ \\
			\hline \hline
			Ours & \textbf{74.40} & \textbf{74.74}  &  \textbf{0.7052}  \\
			3D U-Net & 74.08 & 72.70 &   0.7989 \\
			$\text{SWI}_\text{ATRG}$ & 70.23 &61.84 &   1.413  \\
			\hline
		\end{tabular}
	\end{adjustbox}\vspace{-2mm}
\end{table}

\vspace{-1mm}
\subsubsection{Discussion on Labeling and Visualization Tool}
\label{sec:discussion}
It is interesting to note that in the TubeTK dataset even the ground truth vessel mask does not cover certain vessel continuity, which can be clearly traced on MIPs (such as some yellow circles in the ground truth MIPs in Fig.~\ref{visual1}), since it is very difficult to label all the vessels in corresponding MRA slices in a single slice-by-slice manipulation without referring to MIP and 3D global visualization. However, based on the ground truth MIP labeling slices in the MICRO-MRI datset from our experiments and collaborative evaluations, we can see that such issue is greatly alleviated, since our cerebrovascular labeling and visualization tool is applied to generate (refine) these ground truth vessel labels. In the future, we will further refine the ground truth vessel labels of TubeTK dataset by using our developed labeling tool (such as examples shown in Supplemental Video) under the domain experts' guidance for better public sharing and use.

\begin{figure}[t] 
	\begin{center}
		\includegraphics[width=0.81\linewidth]{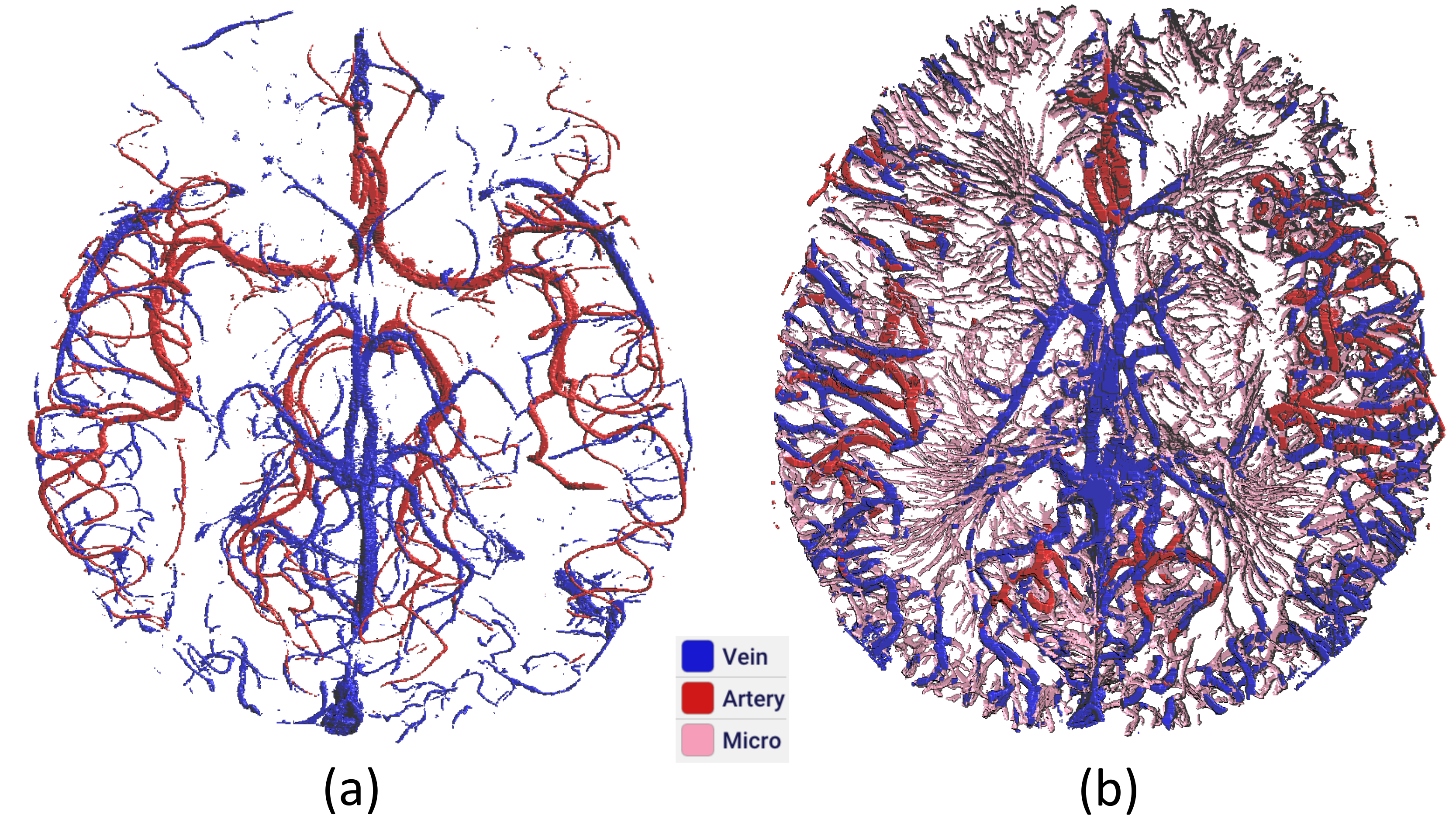}\vspace{-2mm} 
		\caption{Joint 3D visualization of our segmentation results on MICRO-MRI dataset in two different testing cases: (a) Whole midbrain major-level arteries (in red) and veins (in blue). (b) Major-level arteries (in red), major-level veins (in blue), and micro-level vessels (in pink) from slice No. 20 to 40 within midbrain area. Some large pink vessels are also major-level ones which are absent from major-level MRAGs and MRVGs due to the different image acquisitions.} \vspace{-8mm}
    \label{all-fuse}
	\end{center}
\end{figure}

Last but not least, to our knowledge, we are the first to investigate and apply our 3D brain vasculature segmentation to different vessel types and levels, especially the micro-level vessel segmentation; also, it is the first time that the whole brain vessels with different types / levels can be visualized \emph{in-vivo}. Therefore, we have also designed a visualization tool for jointly showing different vasculature systems. Our tool enables the visualization for any combination of different vessel systems in user-defined color and lighting, and support all essential auxiliary interactions such as rotation, translation, scaling, zooming in / out, clipping, etc., for better examination. Fig.~\ref{all-fuse} (a) shows the joint visualization for our prediction results of major-level midbrain arteries and veins; and Fig.~\ref{all-fuse} (b) shows all three vasculature systems aligned together, i.e., major-level midbrain arteries and veins, and micro-level vessels, from MICRO-MRI dataset. Supplemental Video is included for demonstrating the dynamic visualization and interaction in detail.
\vspace{0mm}
\section{Conclusion}
In this work, we have proposed the VC-Net, a deep neural network to extract and visualize high-fidelity 3D cerebrovascular structure from highly sparse and noisy images. VC-Net has three major components, i.e., 3D and 2D dual-domain segmentation streams, 3D-to-2D projection for two-stream design, and 2D-to-3D unprojection for joint embedding operations. By unprojecting the learned multislice composited 2D MIP feature vectors into the 3D volume embedding space, the proposed framework can strengthen the sparse 3D vascular representation by better capturing the small / micro vessels as well as improving the vessel connectivity, which outperforms the state-of-the-art classical and deep learning based methods. In medical practice, this work can be used as the key functions for real-time \emph{in-vivo} segmentation and visualization of sparse and complicated 3D microvascular structure to improve MICRO MRAV diagnosis of vascular disease.

In the future, we will continue to explore research problems related to volume rendering supported 3D exploration and analysis to leverage both the 2D findings and the 3D knowledge and analytics by deep neural networks. We will extend current MIP-based volume rendering (i.e., a special case of volume rendering) into more general volume rendering scenarios, such as X-ray projections, full RGB composition, multi-view MIPs, and flow modeling concepts.
\vspace{-1mm}
\section*{Acknowledgments}
We would like to thank the reviewers for their valuable comments. We are grateful to Yongsheng Chen from Neurology for the early discussion of this work, Pavan K. Jella from Radiology for preparing and collecting the clinical datasets, and Michelle Hua from Cranbrook Schools for pre-processing the datasets and proofreading the paper. This work was partially supported by the NSF under Grant Numbers IIS-1816511, CNS-1647200, OAC-1657364, OAC-1845962, OAC-1910469, the Wayne State University Subaward 4207299A of CNS-1821962, NIH 1R56AG060822-01A1, NIH 1R44HL145826-01A1, ZJNSF LZ16F020002, and NSFC 61972353.

\bibliographystyle{abbrv-doi}

\bibliography{template}

\clearpage

\section*{Supplemental Material}
\beginsupplement
\section{Supplemental Figures}
\begin{figure*}[htb!]
	\begin{center}
		\includegraphics[width=0.83\textwidth]{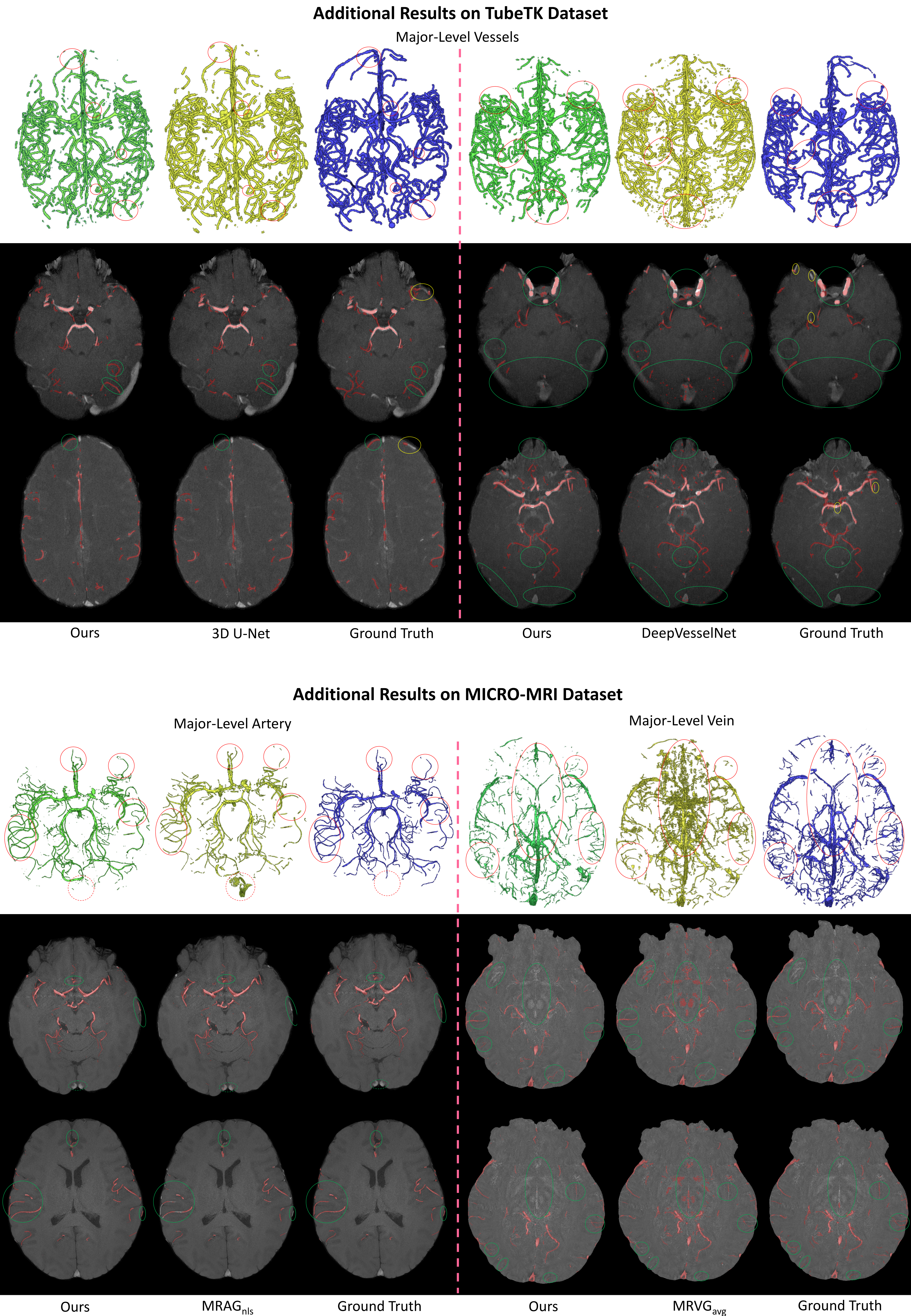}\vspace{-0mm}
		\caption{Additional qualitative results from two datasets (top: TubeTK dataset, bottom: MICRO-MRI major-level vessel dataset): The 3D global vessel segmentations are shown from superior direction. The MIP segmentations are visualized by 5-sliced MRA / MICRO-MRI images, and the corresponding vessel masks in MIPs are marked in semi-transparent red. The highlighted comparison areas are marked in circles. The 3D MRAG / MRVG images from MICRO-MRI dataset only focus on midbrain area and thus have less vessels compared with TubeTK dataset. It is noted that in TubeTK dataset even the ground truth vessel label does not perfectly cover certain vessel continuity, which can be clearly traced on MIPs (such as some yellow circles in the ground truth MIPs), in the corresponding MRA slices. We will further refine the ground truth vessel labels of TubeTK dataset by using our developed labeling tool under the domain experts' guidance in our future work.}
     \label{visual_additional}
	\end{center}
\end{figure*}

\begin{figure*}[htb!]
	\begin{center}
		\includegraphics[width=0.7\textwidth]{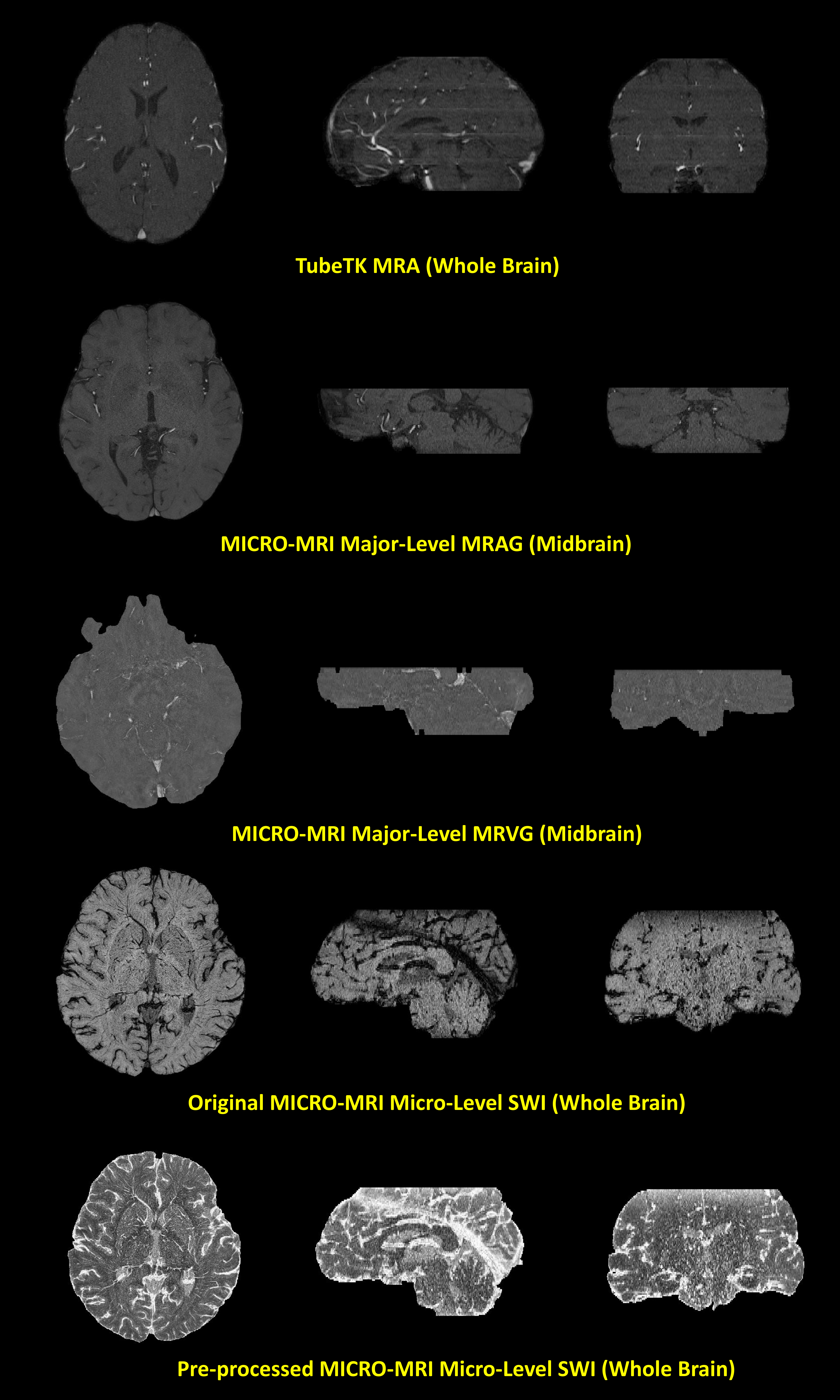}\vspace{-0mm}
		\caption{The different modalities of input image examples from TubeTK MRA and MICRO-MRI datasets in our VC-Net method.}
		\label{input}
	\end{center}
\end{figure*}

Supplemental figures are included for demonstrating additional qualitative results from TubeTK and MICRO-MRI datasets in Fig.~\ref{visual_additional}, and different modalities of input image examples from TubeTK MRA and MICRO-MRI datasets in our VC-Net method in Fig.~\ref{input}.

\section{Additional Quantitative Performance Evaluation}
In order to further demonstrate the effectiveness of our VC-Net (especially 3D-to-2D projection in dual-stream and 2D-to-3D unprojection for joint embedding in our proposed architecture), Tab.~\ref{tab_combination} shows the numerical analyses on some simple combinations of the final results from 3D U-Net and 2D U-Net through average and max operations on the probabilities.
\begin{table}[htb!]
	\caption{Quantitative performance evaluation between different combinations of 3D U-Net and 2D U-Net and our method on TubeTK dataset.}\vspace{-1mm}
	\label{tab_combination}
    \centering
	\begin{adjustbox}{width=0.45\columnwidth,center}
		\begin{tabular}{|c|c|c|c|c|}
			\hline
			Methods / Metrics & Dice (\%) $\uparrow$ \\
			\hline \hline
			2D U-Net & 65.10  \\
            3D U-Net & 71.01  \\
			Average Fusion & 65.15  \\
            Max Fusion & 69.41  \\
            Ours & \textbf{71.81}  \\
			\hline
		\end{tabular}
	\end{adjustbox}
\end{table}

From Tab.~\ref{tab_combination}, we can see our VC-Net overall outperforms both combination methods of the final results of 3D U-Net and 2D U-Net. As shown in Sec. 4.1 of the paper, 2D U-Net performs much worse than a standalone 3D U-Net on each metric. Unlike the 2D composited MIP stream in VC-Net, 2D U-Net itself essentially does not involve any complementary or enhancement information, and the reception field of 2D U-Net is restricted to an isolated 2D slice patch every time and thus lack of the contextual information from the third dimension, which is fatal to the sparse 3D vessel segmentation. Without comprehensive 3D spacing neighborhood, 2D U-Net is more prone to strong noise perturbation (high-intensity true negative) and insensitive to weak vessel signal (low-intensity true positive), as a result, 2D U-Net performs unsatisfactorily even when equipped with more feature embedding channels. Consequently, it may not be an ideal idea to fuse the results from 3D U-Net and 2D U-Net through the simple combinations.

Here, Dice Similarity is provided since it is the most comprehensive and effective indicator / metric to justify the segmentation performance. It measures the intersection over union between the prediction and the ground truth, which comprehensively takes into account all true positive (TP), false negative (FN), as well as false positive (FP). This is also why we (as well as many other research works) select Dice Similarity as the loss function in our VC-Net.

\section{Labeling Refinement and Visualization Tool}
\begin{figure*}[htb!]
	\begin{center}
		\includegraphics[width=0.9\textwidth]{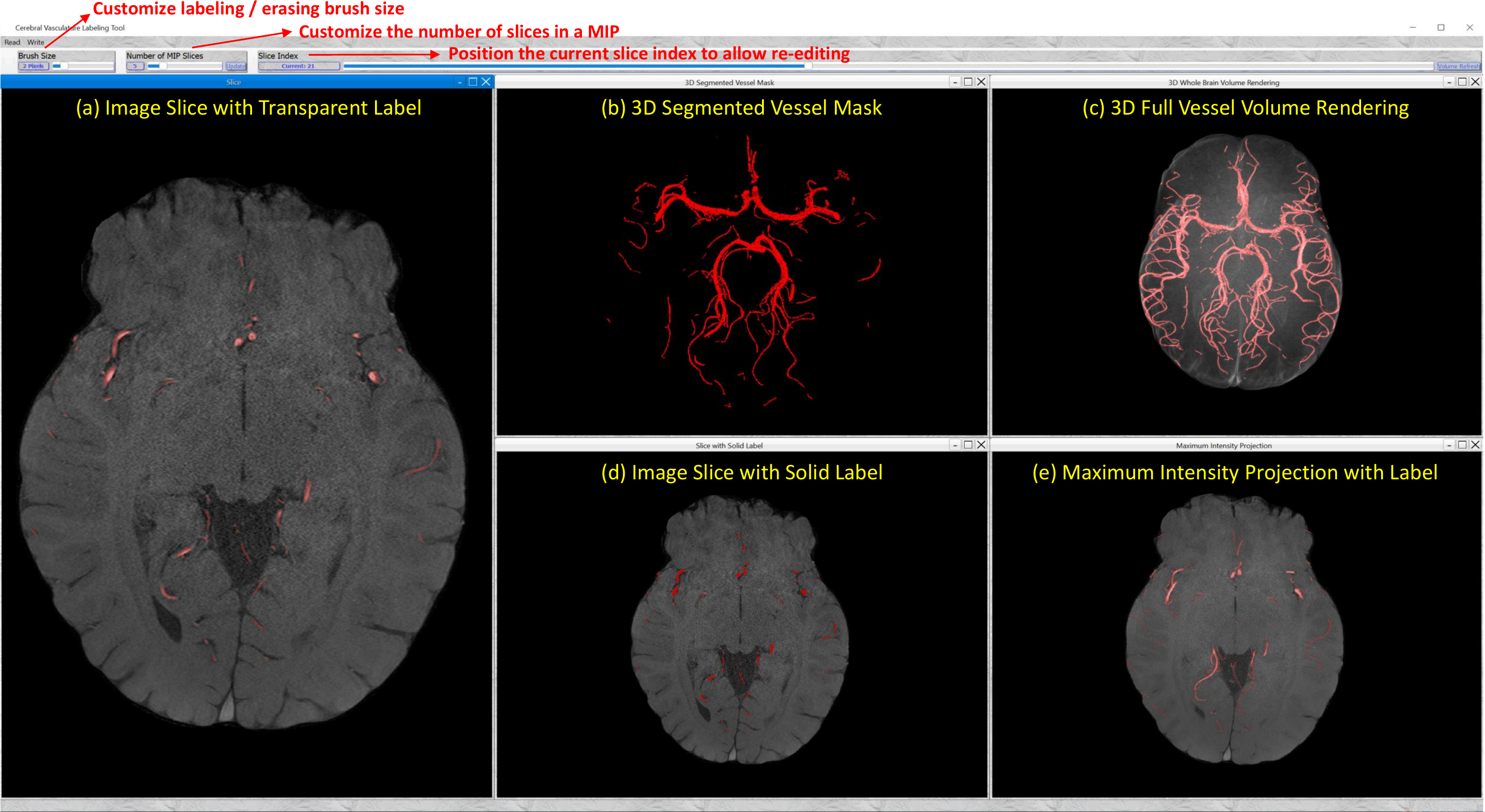}
		\caption{Our developed cerebrovascular labeling and visualization tool (e.g., an example of major-level arterial vessels from MICRO-MRI).}
		\label{tool}
	\end{center}
\end{figure*}

The interface and basic functions of our specifically-designed cerebrovascular labeling and visualization tool are shown in Fig.~\ref{tool}. Our tool enables slice-wise refinement based on the pre-computed vessel labels by $\text{MRAG}_{\text{nls}}$, $\text{MRVG}_{\text{avg}}$, and $\text{SWI}_{\text{ATRG}}$ methods, instead of labeling from scratch manually. The interactive vessel editing is conducted in the current image slice window, e.g., manually labeling / erasing brush, automatically labeling connected components by flood-fill method as shown in Fig.~\ref{tool} (a). The slice under editing is simultaneously visualized in solid red for a clearer examination in Fig.~\ref{tool} (d). Unlike the operation in most of the general-purpose labeling / segmentation softwares in which the current labeling (2D) slice is usually isolated from its (3D) context and thus lacks the crucial reference, the vessel labeling in our developed tool is comprehensively assisted and guided by the following specifically-desired functions: (1) simultaneously updated 3D vasculature system from the beginning to the current slices with several interactions, such as rotation and zooming in / out, to check the cross-plane 3D vessel connectivity (Fig.~\ref{tool} b); (2) synchronized brain vessel volume rendering to trace the overall segmented vasculature system (Fig.~\ref{tool} c); (3) adaptive MIP labeling display (with user-defined number of projection slices) that enables users to evaluate the contextual slices to strengthen the vessel connectivity and rule out noise (Fig.~\ref{tool} e). Our tool can greatly facilitate the continuous slice-wise labeling and reduce the labeling ambiguity in some challenging areas of the micro-cerebrovascular structure, which have been extensively tested and evaluated by our collaborative domain experts. Supplemental Video is included to demonstrate the dynamic visualization and interaction in detail.

\section{Supplemental Video}
Supplemental video is included to demonstrate the joint 3D visualization of the major-level and micro-level vessels in the midbrain and the whole brain on MICRO-MRI dataset; as well as the dynamic visualization and interaction of our developed cerebrovascular labeling and visualization tool.

\end{document}